\documentclass[3p,12pt,preprint]{elsarticle}

\usepackage{lineno,hyperref}
\modulolinenumbers[1]

\newcommand{\bsub}{\begin{subequations}}
\newcommand{\esub}{\end{subequations}}

\newcommand{\ga}{\gamma}

\newcommand{\beq}{\begin{equation}}
\newcommand{\eeq}{\end{equation}}
\newcommand{\bsubeq}{\begin{subequations}}
\newcommand{\esubeq}{\end{subequations}}
\newcommand{\beqn}{\begin{eqnarray}}
\newcommand{\eeqn}{\end{eqnarray}}
\newcommand{\fr}{\frac}
\newcommand{\lb}{\label}
\newcommand{\er}{\eqref}

\usepackage{hyperref}
\hypersetup{colorlinks,allcolors=blue}
\usepackage{amsmath}
\usepackage{multirow}
\usepackage{amsfonts,amssymb}
\usepackage{graphicx}
\usepackage{subfigure}
\usepackage[rel]{overpic}

\graphicspath{{figures/}}

\begin{document}


\begin{frontmatter}

\title{\textbf{A new similarity law for transonic-supersonic flow}}


\author{Luoqin Liu\corref{cor1}}
\cortext[cor1]{Corresponding author}
\ead{luoqin.liu@utwente.nl}


\address{Physics of Fluids Group, Max Planck Center Twente for Complex Fluid Dynamics, University of Twente, 7500 AE Enschede, The Netherlands}


\begin{abstract}
How to determine accurately and efficiently the aerodynamic forces of the aircraft in high-speed flow is one of great challenges in modern aerodynamics. In this Letter we propose a new similarity law for steady transonic-supersonic flow over thin bodies. The new similarity law is based on the local Mach number frozen principle. It depends on both the specific heat ratio and the free-stream Mach number. The new similarity law enables one to determine the lift and drag coefficients of the aircraft from that of a reference state which is more reachable. The validity of the new similarity law has been confirmed by the excellent agreement with numerical simulations of both two-dimensional airfoil flows and three-dimensional wing flows. 
\end{abstract}

\begin{keyword}
Similarity law, transonic flow, supersonic flow, lift, drag
\end{keyword}

\end{frontmatter}



{\it Introduction.} 
The similarity laws of aerodynamics were developed starting in the twenties of the last century \citep{Ackeret1925, Glauert1928, Prandtl1930, Tsien1939, Karman1947b}. They were usually obtained by solving the steady full potential equation, either through linearization approach \citep{Ackeret1925, Glauert1928, Prandtl1930} or hodograph transformation \citep{Tsien1939}. However, as the simplified full potential equation has different mathematical properties at subsonic, transonic, and supersonic flows, these similarity laws are valid only over a certain range of Mach number and cannot transform smoothly from one to another. In this Letter we derive a new similarity law for steady high-speed flow over thin bodies. Instead of solving the full potential equation, the derivation is based on the local Mach-number frozen principle. Therefore, the new similarity law holds for both transonic and supersonic flows, provided that subsonic and supersonic regimes coexist on the aircraft surfaces. The validity of the proposed similarity law is confirmed by the numerical simulations of steady flows over a two-dimensional airfoil and a three-dimensional wing.

{\it Similarity law.}
Consider the steady flow of perfect gas over a thin airfoil. The pressure coefficient $C_p$ on the airfoil surface is defined by
\beq 
C_p = \fr{p-p_\infty}{\fr{1}{2} \rho_\infty U_\infty^2} = \fr{2}{\gamma M_\infty^2} \left( \fr{p}{p_\infty} - 1 \right),
\eeq 
where the subscript $\infty$ denotes the value at infinity of upstream, $p$ is the pressure, $\rho$ is the density, $U$ is the velocity magnitude, $\gamma$ is the specific heat ratio, and $M$ is the Mach number. 
The pressure $p$ is related to the stagnation pressure $p_0$, namely
\beq 
\fr{p_0}{p} = \left( 1 + \fr{\gamma-1}{2} M^2 \right)^{ \fr{\gamma}{\gamma-1}}.
\eeq 
For isentropic flows, the stagnation pressure on the body surface is the same as that at far upstream. 
For flows with only weak shocks, the variance of stagnation pressure is only a third-order small term \citep{Liepmann1957}. These two facts enable us to neglect the variance of stagnation pressure. Thus, 
\begin{equation}\label{eq.cp}
  C_p = \fr{2}{\ga M_\infty^2} \left\{ \left[ \fr{2+(\ga-1) M_\infty^2}{2+(\ga-1)M^2} \right]^{ \fr{\ga}{\ga-1}} -1 \right\}.
\end{equation}
To calculate $\textrm{d} C_p / \textrm{d} M_\infty$, we rewrite \er{eq.cp} as
\beq \lb{eq.cp-1}
\ln \left( 1 + \fr{\gamma M_\infty^2}{2} C_p \right) = \fr{\gamma}{\gamma-1} \ln \left[ \fr{2+(\ga-1) M_\infty^2}{2+(\ga-1)M^2} \right].
\eeq
Differentiating equation \er{eq.cp-1} with respect to $M_\infty$, there is
\beq\lb{eq.dCp-dMinfty-0} 
\begin{split}
& \fr{\textrm{d} C_p}{\textrm{d} M_\infty} = \fr{2}{M_\infty} \fr{ 2 - (2 - M_\infty^2) C_p }{ 2+(\ga-1) M_\infty^2 } 
 -\left( C_p + \fr{2}{\gamma M_\infty^2} \right) \fr{2 \gamma M}{ 2+(\ga-1) M_\infty^2 } \fr{\textrm{d} M}{\textrm{d} M_\infty}.
\end{split}
\eeq 
On the one hand, the local Mach-number frozen principle states that $M$ becomes stationary as $M_\infty$ passes through unity \citep{Liepmann1957, Ramm1990},
\begin{equation}\label{Mach}
  \left( \fr{\textrm{d} M}{\textrm{d} M_\infty} \right)_{M_\infty = 1} =0.
\end{equation}
This fact enables us to safely neglect the second term in the right hand side of equation \er{eq.dCp-dMinfty-0}, provided that $M_\infty$ is large enough such that supersonic and subsonic zones coexist on the body surface. 
On the other hand, to simplify the integration procedure, we set $M_\infty=1$ in the first term of equation \er{eq.dCp-dMinfty-0}. 
Thus, equation \er{eq.dCp-dMinfty-0} can be approximated as
\beq\lb{eq.dCp-dMinfty-2} 
  \fr{\textrm{d} C_p}{\textrm{d} M_\infty} = \fr{2}{\ga+1} \left( 2 - C_p \right).
\eeq
This is a first-order ordinary differential equation for the pressure coefficient. 
Note that equation~\er{eq.dCp-dMinfty-2} is exactly valid at $M_\infty=1$ \citep{Liepmann1957, Ramm1990}. But now we presumably that it is also approximately valid in the vicinity of $M_\infty$.

The lift coefficient is calculated as 
\beq \lb{eq.cl-1}
C_l (M_\infty) = \fr{1}{c} \int_{0}^{c} (  C_{p, L} - C_{p, U} ) \textrm{d} x,
\eeq 
where $c$ is the chord length, and the subscripts $L$ and $U$ denote the lower and upper airfoil surfaces. 
From the mean value theorem, equation \er{eq.cl-1} can be integrated as 
\beq \lb{eq.cl-2}
C_l (M_\infty) = C_p(M_L; M_\infty) - C_p(M_U; M_\infty),
\eeq 
where $M_L$ and $M_U$ are the characteristic Mach numbers on the lower and upper airfoil surfaces, respectively. 
Combining equations \er{eq.dCp-dMinfty-2} and \er{eq.cl-2}, we get a first-order ordinary differential equation for the lift coefficient, 
\beq \lb{dcl-dMinfty-3}
\fr{\textrm{d} C_l}{\textrm{d} M_\infty} = - \fr{2}{\ga+1} C_l.
\eeq
The solution of equation \er{dcl-dMinfty-3} is
\beq \label{Cl-Minfty}
\fr{C_l}{C_{l, \sqrt 2}} = \exp \left[\fr{2}{\ga+1} \left( \sqrt 2 - M_\infty \right) \right], 
\eeq
where $C_{l, \sqrt 2}$ is the lift coefficient of the aircraft at the reference state $M_\infty = \sqrt 2$. Note that in thin airfoil theory $C_{l, \sqrt 2}= 4 \alpha_0$, where $\alpha_0$ is incident angle of the free stream \citep{Liepmann1957}.

Following \citet{Ramm1990}, the drag coefficient $C_d$ can be written as
\beq\lb{eq.cd-0}
C_d (M_\infty) = \fr{1}{2h} \int_{-h}^{h} \alpha (y) C_p(M(y); M_\infty) \textrm{d} y,
\eeq 
where $h$ is the half-thickness of the airfoil and $\alpha$ is the local angle of attack. From the mean value theorem, equation \er{eq.cd-0} can be writen as
\beq\lb{eq.cd-1}
C_d (M_\infty) = \bar \alpha C_p(M_d; M_\infty),
\eeq 
where $M_d$ is a characteristic Mach number related to the drag coefficient, and  
\beq 
\bar \alpha = \fr{1}{2h} \int_{-h}^{h} \alpha (y) \textrm{d} y
\eeq 
is the mean value of the angle of attack. 
We emphasis that for shock-free flow there is $\bar \alpha = \alpha_0$. For shock flow, however, both shock wave induced flow separation and shock wave increased wake width will increase the drag, and thus $\bar \alpha \ne \alpha_0$. 
In general, $\bar \alpha$ should depend on both $M_\infty$ and $\alpha_0$. For simplicity, we assume it is independent of $M_\infty$. 
Then, from equations \er{eq.dCp-dMinfty-2} and \er{eq.cd-1} we obtain a first-order ordinary differential equation for the drag coefficient, 
\beq \lb{eq.dcd-dMinfty-1}
\fr{\textrm{d} C_{d} }{\textrm{d} M_\infty} = \fr{2}{\ga+1} \left( 2\bar \alpha - C_{d} \right). 
\eeq
The solution of equation \er{eq.dcd-dMinfty-1} is
\begin{equation}\label{eq.cd-Minfty}
\fr{C_d-2\bar \alpha}{C_{d, \sqrt 2} -2\bar \alpha} = \exp \left[\fr{2}{\ga+1}(\sqrt 2-M_\infty) \right],
\end{equation}
where $C_{d, \sqrt 2}$ is the drag coefficient of the aircraft at the reference state $M\infty = \sqrt 2$. Note that in thin airfoil theory $C_{d, \sqrt 2}= 4 \alpha_0^2$ \citep{Liepmann1957}.

We remark that, the new similarity law for the lift coefficient \er{Cl-Minfty} and the drag coefficient \er{eq.cd-Minfty} has a unique feature compared to existing similarity laws \citep{Ackeret1925, Glauert1928, Prandtl1930, Tsien1939, Karman1947b}. That is, the new similarity law depends on $M_\infty^1$ while the existing ones depend on $M_\infty^2$. Similar to the existing transonic similarity laws \citep[e.g.,][]{Karman1947b}, the new similarity law also depends on the specific heat ratio $\gamma$. Besides, the new similarity law for the drag coefficient \er{eq.cd-Minfty} has considered the effects of shock-induced flow separation and shock-increased wake width. 
In addition, although the new similarity law of equations \er{Cl-Minfty} and \er{eq.cd-Minfty} is derived for two-dimensional thin airfoils, it is also valid for three-dimensional thin wings. This is because the essential assumption of the derivation is based on the local Mach number frozen principle, which is independent of the dimensionality of the aircraft and the space.

{\it Numerical validation.}
To verify the correctness of the new similarity law given by equations~\er{Cl-Minfty} and \er{eq.cd-Minfty} for transonic-supersonic flow, comparisons with the numerical results are made for steady viscous and compressible flows over both a two-dimensional airfoil and a three-dimensional wing.
In two dimensions, \citet{Liu2015} performed a series of Reynolds-averaged Navier-Stokes simulations of the flow over an RAE-2822 airfoil. In these simulations, the Reynolds number is $Re=6.5\times10^6$, and two angles of attack are considered, namely $\alpha_0=2.31^\circ$ and $5.00^\circ$. The simulations were performed by the OpenCFD-EC2D-1.5.4 program developed by Professor Xinliang Li of the Chinese
Academy of Sciences. More details of the simulations can be found in \citet{Liu2015}. 
In three dimensions, \citet{Vinh2015} performed a series of large eddy simulations of the flow over an isolate wing. In these simulations, the Reynolds number is about $Re=5.0\times10^5$ and the angle of attack is $\alpha_0=20.0^\circ$. The simulations were performed with the commercial software package ANSYS. More details of the simulations can be found in \citet{Vinh2015}.

\begin{figure}[!tb]
\centering
\includegraphics[width=0.65\textwidth]{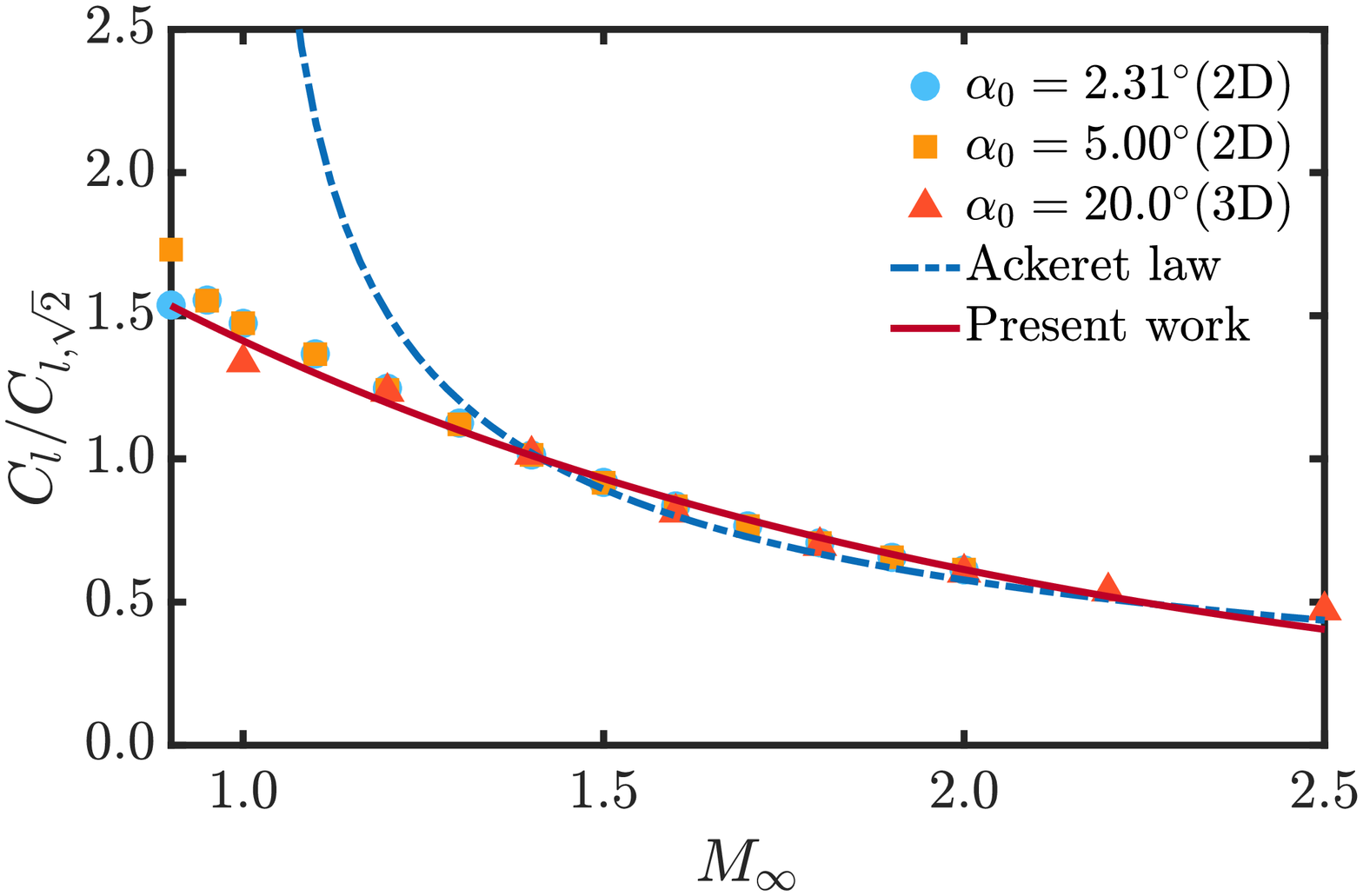}
\caption{Mach-number dependence of the lift coefficient of a two-dimensional airfoil and a three-dimensional wing. Filled circles: numerical results with $\alpha_0=2.31^\circ$ performed by \citet{Liu2015}; filled squares: numerical results with $\alpha_0=5.00^\circ$ performed by \citet{Liu2015}; filled triangles: numerical results with $\alpha_0=20.0^\circ$ performed by \citet{Vinh2015}; dashed-dotted line: the prediction of the Ackeret law given by equation~\er{eq.Ackeret}; solid line: the prediction of the new similarity law given by equation \er{Cl-Minfty}. }
\label{fig.L2-5}
\end{figure}

Figure~\ref{fig.L2-5} shows the comparison of the new similarity law for the lift coefficient given by equation~\er{Cl-Minfty} and the numerical simulations performed by \citet{Liu2015} and \citet{Vinh2015}. Also shown is the theoretical prediction of the Ackeret law \citep{Ackeret1925}, 
\beq \lb{eq.Ackeret}
\fr{C_l}{C_{l, \sqrt 2}} = 
\fr{C_d}{C_{d, \sqrt 2}} = \fr{1}{\sqrt{M_\infty^2-1}}. 
\eeq 
Note that the theoretical results obtained from both the new similarity law for the lift coefficient given by equation~\er{Cl-Minfty} and the Ackeret law given by equation~\er{eq.Ackeret} agree excellently with the numerical results at the supersonic regime $M_\infty \ge 1.4$. 
But at the transonic regime $0.9 \le M_\infty \le 1.3$, only the new similarity law agrees well with the numerical results. This result shows the superiority of the new similarity law over the Ackeret law.

\begin{figure}[!tb]
\centering
\includegraphics[width=0.65\textwidth]{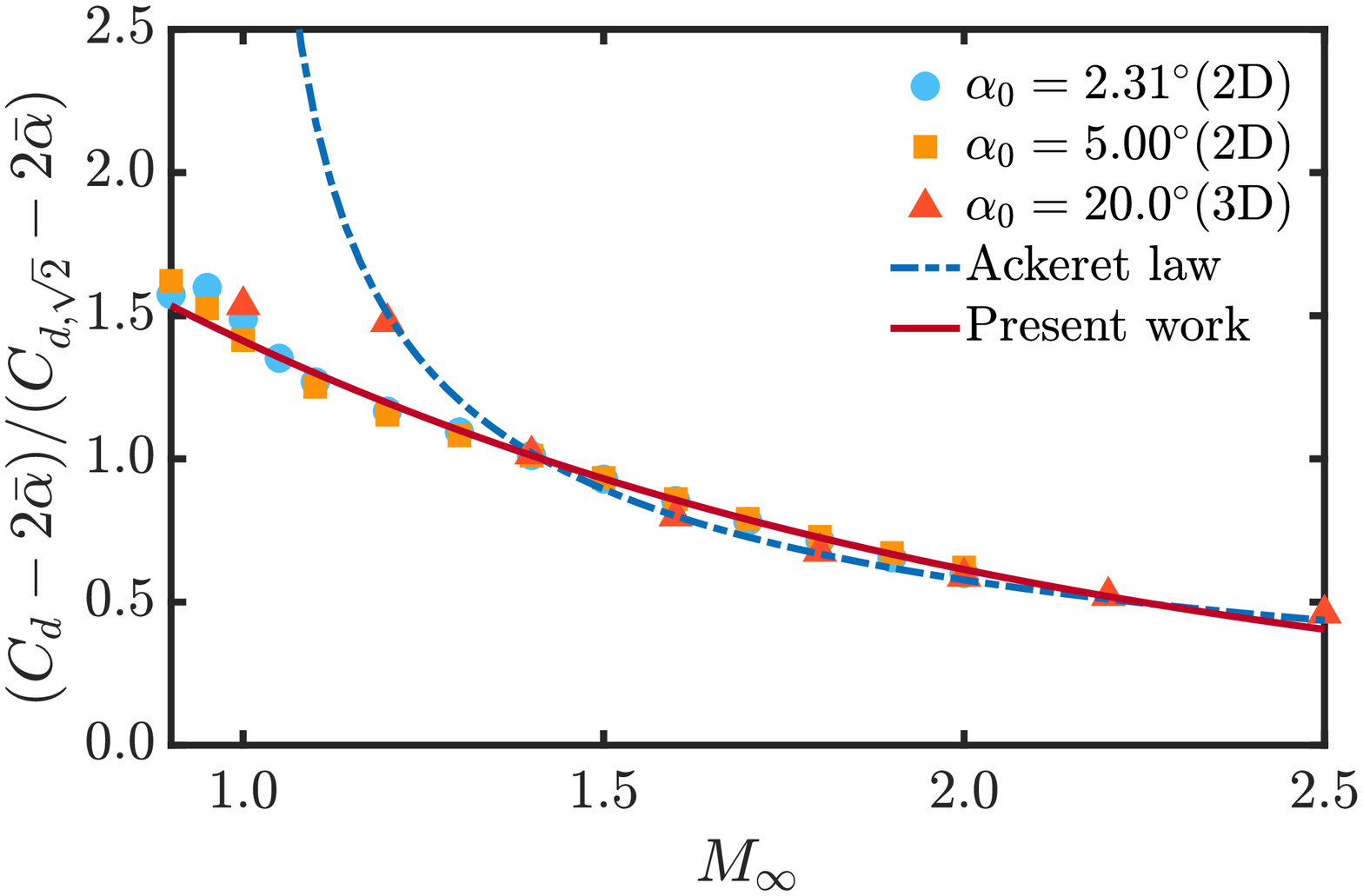}
\caption{Mach-number dependence of the drag coefficient of a two-dimensional airfoil and a three-dimensional wing. Filled circles: numerical results with $\alpha_0=2.31^\circ$ performed by \citet{Liu2015}; filled squares: numerical results with $\alpha_0=5.00^\circ$ performed by \citet{Liu2015}; filled triangles: numerical results with $\alpha_0=20.0^\circ$ performed by \citet{Vinh2015}; dashed-dotted line: the prediction of the Ackeret law given by equation~\er{eq.Ackeret}; solid line: the prediction of the new similarity law given by equation \er{eq.cd-Minfty} with $\bar \alpha = 0.03$. }
\label{fig.d2-5}
\end{figure}

Figure~\ref{fig.d2-5} shows the comparison of the new similarity law for the drag coefficient given by equation~\er{eq.cd-Minfty} and the numerical simulation performed by \citet{Liu2015} and \citet{Vinh2015}. Also shown is the theoretical prediction of the Ackeret lawgiven by equation~\er{eq.Ackeret}.
Note that the theoretical results obtained from both the new similarity law for the drag coefficient given by equation \er{eq.cd-Minfty} with $\bar \alpha = 0.03$ and the Ackeret law given by equation~\er{eq.Ackeret} agree excellently with the numerical results at the supersonic regime $M_\infty \ge 1.4$. 
But at the transonic regime $0.9 \le M_\infty \le 1.3$, only the new similarity law agrees well with the numerical results. 
Note that the normalized drag coefficient $(C_d-2\bar \alpha)/(C_{d, \sqrt 2}-2 \bar \alpha)$ of the three-dimensional wing at $M_\infty=1.2$ falls accidentally on the curve of the Ackeret law. The reason for this isolate point is not known yet since we don't know the details of the simulation. Therefore, more careful simulations of three-dimensional thin bodies are helpful to further evaluate the performance of the new similarity law for the drag coefficient. Nevertheless, since all other points of the three-dimensional wing  (including the sonic point at $M_\infty=1$) and all points of the two-dimensional airfoil nearly collapse to the curve of the new similarity law, we confirm once more the superiority of the new similarity law over the Ackeret law.

{\it Conclusions.}
In this Letter we propose a new similarity law for high-speed flow over thin bodies. The new similarity law is based on the local Mach number frozen principle and thus is valid for both transonic and supersonic flows. In contrast to existing similarity laws which depend on the square of free-stream Mach number $M_\infty^2$, the new similarity law depends on $M_\infty^1$. Similar to the existing transonic similarity laws, the new similarity law also depends on the specific heat ratio $\gamma$. 
The new similarity law enables one to determine the lift and drag coefficients of the aircraft in steady transonic-supersonic flow from that of a reference state. The validity of the new similarity law has been confirmed by the excellent agreement with numerical simulations of the flow over both a two-dimensional airfoil and a three-dimensional wing.

\section*{Acknowledgments}
\addcontentsline{toc}{section}{Acknowledgements}

The author is grateful to Prof. Jiezhi Wu for insightful discussions. The author acknowledges Dr. Shufan Zou for providing the literature of wing simulations.

\addcontentsline{toc}{section}{References}
\bibliographystyle{model5-names}\biboptions{authoryear}
\bibliography{mybibfile}

\end{document}